\documentclass[journal]{IEEEtran}
\usepackage{graphicx}
\ifCLASSINFOpdf
  
\else

\fi

\usepackage[english]{babel}
\usepackage[dvipsnames,svgnames,x11names]{xcolor}
\usepackage[markup=underlined]{changes}

\usepackage[cmex10]{amsmath}

\begin{document}

\title{
  Analysis of Noise in Current Mirrors with memristive Device\\
  }

\author{Nazerke~Kulmukhanova and  Irina Dolzhikova,~\IEEEmembership{ Student Member,~IEEE,}\\
   
\IEEEauthorblockA{Electrical and Computer Engineering Department, Nazarbayev University, Astana, Kazakhstan
}
}

\maketitle

\IEEEpeerreviewmaketitle
\begin{abstract}
This work presents an analysis of noise in a cascode current mirror with CMOS-memristive device done by comparison with the basic current mirror. The analysis is completed based on THD for different frequency and channel length values by means of computer-aided design. AC and DC analyses are presented for both balanced and unbalanced current mirrors. While the change in the channel length has similar effect in both circuits, memristor in a circuit decreases noise significantly at high frequencies.   
\end{abstract}


\section{Introduction}

\IEEEPARstart {}
Current mirrors (CM) are widely used circuits that allow to copy a reference current level and, thus, control the current through the other device. It is widely used in operational amplifiers and integrated circuits. Current mirror eliminates a need to establish a current source with specific characteristics multiple times which is a very complex procedure. Instead, similar result is achieved by using only two transistors. It is also used in most operational amplifier circuits, and is among the building blocks of the integrated circuits. The simplest current mirror consists of two transistors as presented in the Fig. 1. It can be further improved to enhance the performance. Thus, there are a lot of variations of current mirrors. The most popular ones are Wilson and cascode current mirror, which are presented in the Fig. 2. They provide a much better performance than a basic CM with minimum increase in complexity. 

\begin{figure}  [!ht] 
\begin{center}  
\includegraphics[scale=0.4]{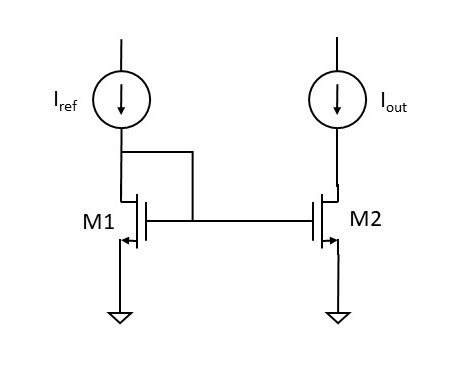} 
\caption{\small \sl Basic CM}
\end{center}     
\end{figure}

\begin{figure}  [!ht] 
\begin{center}  
\includegraphics[scale=0.3]{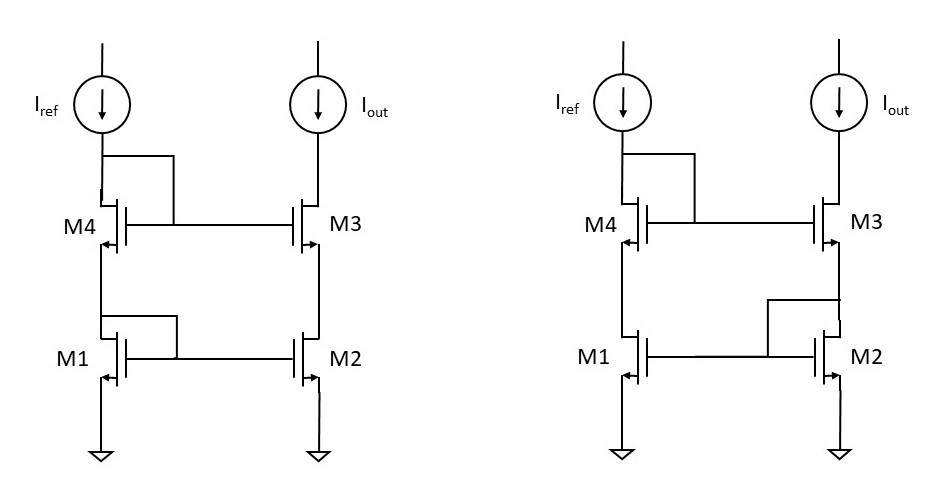} 
\caption{\small \sl Cascode (left) and Wilson (right) CMs}
\end{center}     
\end{figure}

Main noise types in current mirrors include Flicker, Burst, Thermal and Shot noises. According to Bilotti and Mariani \cite{Bilotti}, Cascode, Wilson CMs and their main variations demonstrate similar noise tolerance with slight discrepancies at low and high frequencies. In particular, Cascode CM exhibits a consistent performance throughout the whole range of frequencies, while Wilson CM performed better at high frequencies. Another parameter that affects the noise in CMs is channel length. The same study \cite{Bilotti} suggests that when it is lower than 1.6 $\mu$m Wilson CM is more tolerant to noise than other configurations, but there is no notable difference for greater channel length values.

Memristor is a promising device that is gaining popularity as more and more applications of its memory effect are being discovered \cite{bookJames}. The main working principle is that the oxide level in it can change depending on a current flowing through. When current is no longer applied its oxide layer remains on that particular level and resistance stays fixed at a new value. This effect can be used for non-volatile memory and neuromorphic circuits \cite{irmanova2018neuron}. In software memristor models are developed mostly by utilizing a concept of parallel switching channels that can have on and off states \cite{Meta}, \cite{irmanova2017multi}. Depending on an applied voltage levels, channels switch ON or OFF and change the overall conductance of the device. This switch-based conductance change is represented as follows in Eq. 1.
\begin{equation}\label{conductance}
    G=\frac{X}{R_{ON}}+\frac{1-X}{R_{OFF}}
\end{equation}

Where X is the number of metastable switches in ON state, which can be calculated by:  

\begin{equation}\label{conductance}
    X=\frac{R_{ON}(R_{init}-R_{OFF})}{R_{init}(R_{ON}-R_{OFF})}
\end{equation}

Thus, oxide level in memristor is represented by a number of conducting channels in the model. This method is described in \cite{Yakopcic} and provides a suitable simulation model of memrsitor. However, memristor models available do not take into account any noise types. Nevertheless, noise in memristor was found to be equivalent to the resistor noise and, thus, have much smaller effect on a circuit than the noise in transistors, and is of no interest.

In this paper, we provide a noise analysis of the cascode current mirror, and investigate the dependence between the noise level and frequency, transistor channel length, memristor resistance. Also, a practical case of a circuit with unbalanced values of channel length and memristor resistance is evaluated.

\section{Small-signal model Analysis}
\subsection{Basic Current Mirror}
For a basic transistor-based current mirror as in Fig. 1, a small-signal model will be as presented in the Fig. 3. Main equations describing its operation are as follows:

\begin{align*}
  v_1 = h_{11}i_{1} + h_{12}v_{2}\\
  i_2 = h_{21}i_{1} + h_{22}v_{2}
\end{align*}
h-parameters can be found from the model by setting corresponding voltage and current values to 0;

\begin{align*}
  h_{11} &= \frac{v_2}{i_2} = \frac{1}{g_{m1} + g_{\pi2}} = \frac{1}{g_{m1}(1+\frac{n}{\beta_{o2}})} \cong \frac{1}{g_{m1}} \\
  h_{12} &= 0\\
  h_{21} &= \frac{i_2}{i_1} = \frac{g_{m2} + r_{\pi2}}{1+g_{m1} r_{\pi2}} = \frac{\beta_{o2}}{1+\frac{g_{m1}}{g_{m2}}\beta_{o2}} \cong \frac{g_{m2}}{g_{m1}} \cong \frac{I_{C2}}{I_{C1}} \cong n \\
  h_{22} &= \frac{i_2}{v_2} = \frac{1}{r_{o2}}\\
\end{align*}
For MOSFET specifically the result will be: 

\begin{align*}
  h_{11} &= \frac{1}{g_{m1}} \\
  h_{12} &= 0\\
  h_{21} &=  \frac{g_{m2}}{g_{m1}} \cong\frac{(\frac{W}{L})_2}{(\frac{W}{L})_1} \cong n \\
  h_{22} &=  \frac{1}{r_{o2}}\\
\end{align*}
From the equations above, the values of input and output resistances will be $\frac{1}{g_{m1}} $ and $\frac{1}{r_{o2}}$ respectively. It can be seen that the output/input ratio is largely determined by geometry of transistors along with the output voltage level.

\begin{figure}  [!ht] 
\begin{center}  
\includegraphics[scale=0.3]{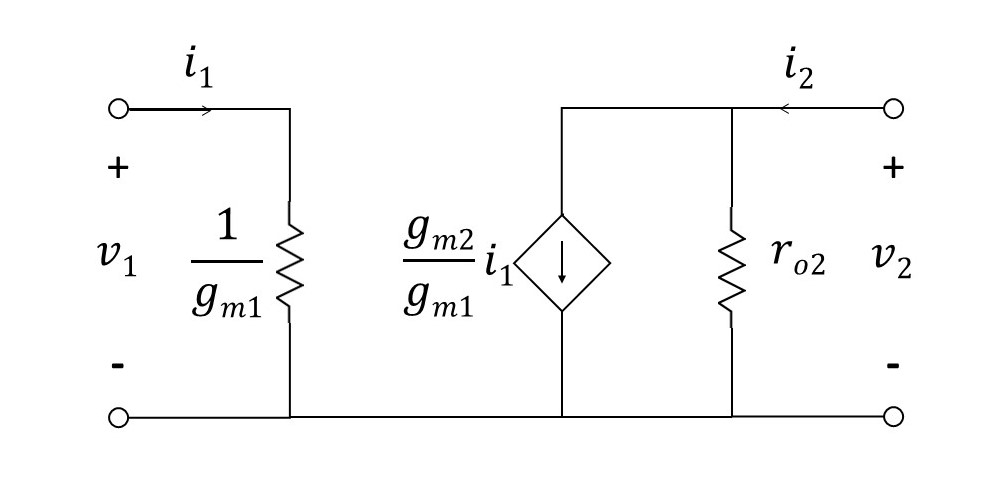} 
\caption{\small \sl Basic Current Mirror small-signal model}
\end{center}     
\end{figure}

In this model many transistor parameters are assumed to be matching, thus not affecting the performance. However, in more realistic model parameters such as oxide layer capacitance, offset threshold and channel length modulation come into picture. 

\subsection{Current Mirror with Memristor}
\begin{figure}  [!ht] 
\begin{center}  
\includegraphics[scale=0.4]{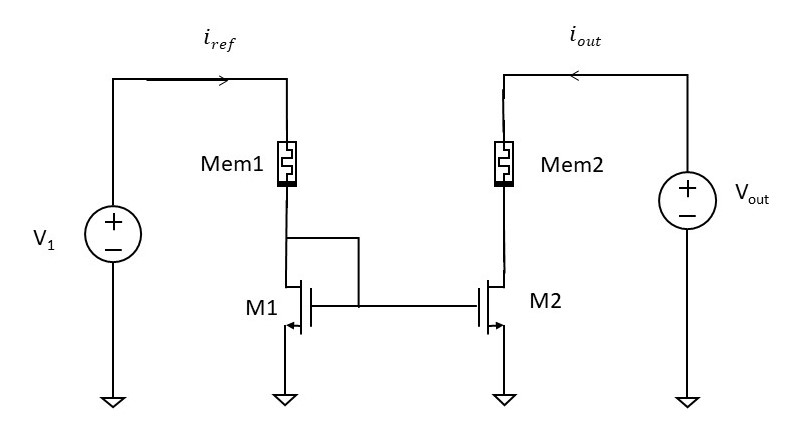} 
\caption{\small \sl The CM with memristors}
\end{center}     
\end{figure}
In the cascode CM two transistors is substituted by memristors. This causes a circuit to be simplified into a basic current mirror with, essentially, non-linear resistors as in Fig. 4. Memristors in the small-signal model are represented as additional resistances in series with existing ones. Thus, the overall operation of the circuit will not change. Input and output resistances become as follows:
\begin{align*}
  r_{in} &= \frac{1}{g_{m1}}+r_{mem1} \\
  r_{out} &=  \frac{1}{r_{o2}+r_{mem2}}\\
\end{align*}
Where $R_{mem1}$ and $R_{mem2}$ are memristor resistance values at input and output sides of current mirrors respectively. When these memristors are identical their resistance expressions are the same. Mathematically these resistances can be expressed through a state variable {\it{w(t)}} with range from 0 to 1 as presented by Z. Biolek \cite{Biolek}:
\begin{align*}
&V(t)=[R_{ON}\frac{w(t)}{D}+R_{OFF}(1-\frac{w(t)}{D})]I(t)\\
&v_D=\frac{dw(t)}{dt}=\frac{\mu_D R_{ON}}{D}I(t)  \\
 &w(t)=\frac{\mu_D R_{ON}}{D}q(t)\\
  \end{align*}
  Where {$\it{v_D}$} is a drift velocity of oxygen deficiencies. A windowing function can be added to ensure that the value of {\it{w(t)}} does not exceed the range from 0 to 1. Then, by equating ${\it{\frac{w(t)}{D}}}$ to {\it{x(t)}}:
  \begin{align*}
 x(t)&=\frac{w(t)}{D}\\
 \frac{dx}{dt} &=  \frac{\mu_D R_{ON}}{D^2}I(t)F(x(t))\\
\end{align*}
a windowing function will be 
 \begin{align*}
F(x(t)) = 1 - (2x(t)-1)^2p
\end{align*}
When value of the state variable is 0 memristor is operating at the lowest resistance $R_{ON}$, and 1 stands for the maximum resistance state $R_{OFF}$. Parameter p defines a sharpness of a window boundary - lower values will provide a smoother transition, while higher values will give an almost rectangular function. 
This mathematical representation was used in memristor Spice model proposed by HP lab \cite{Yakopcic}, and all other models are largely based on it.

\section{Simulation Parameters}

In the simulation of the circuit presented in the Fig. 4 the BSIM 3.1 MOSFET model provided by MOSIS \cite{Model} is used. The Length of the channel is 180nm. A memristor model added in the circuit was developed by Knowm organization \cite{Meta}. Its initial major parameters are $R_{on}$  500$\Omega$, $R_{off}$  1500$\Omega$ and threshold voltage 0.27V.

For an AC analysis because the .noise command in LTSPice does not take into account the effect of memristor in the circuit, a Fourier coefficients are evaluated instead. A comparison is made on the basis of total harmonic distortion. A major problem of the Fourier analysis is that its accuracy is heavily dependent on the resolution of a data sample and its size. However, eliminating this problem by setting a resolution and length of the simulation high is not feasible as it takes too long for the software to evaluate it. So, this calculation error is instead kept constant throughout the simulation by providing the same number of evaluation points and periods for estimation. Thus, for different frequencies appropriate total time and maximum time step are set.

For the DC analysis cases of imbalanced transistor lengths and imbalanced memristor resistances are considered. The channel lengths of both transistors M1 and M2, from the Fig. 4, are varied to observe the dependence between the current and channel length difference.  The parameter LINT in the model file was changed for that \cite{Manual}.For the case of imbalanced memristor resistance values, the memristor in series with the output is varied. The values considered are in the range of 10\% from the original value.

\section{Results}
\subsection{Frequency Dependence}
In the simulation THD for Six different frequencies from 1Hz to 10GHz with logarithmic step of 100 in circuits with and without memristor was observed. The circuit without memristor shows approximately constant noise for the whole range with abrupt fall at 0.1GHz which is caused by a decrease in noise of the main transistor and not yet coming into picture effect of the other transistor. In contrast, the circuit with memristor exhibits a clear decline in noise at higher frequencies. In particular, it becomes 0.03\% when frequency is lower than 10KHz, 0.009\% at 100MHz and 0.000122\% at 10GHz as can be seen in the Fig. 6. 

\begin{figure}  [!ht] 
\begin{center}  
\includegraphics[scale=0.4]{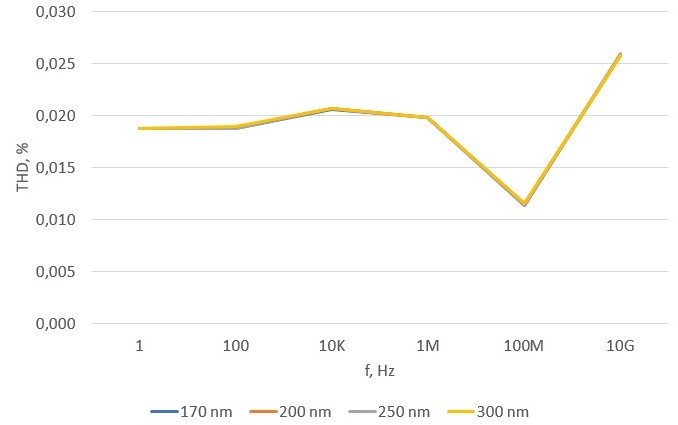} 
\caption{\small \sl THD in a CM without memristor for different channel lengths}
\label{noise_basic} 
\end{center}     
\end{figure}

\begin{figure}  [!ht] 
\begin{center}  
\includegraphics[scale=0.4]{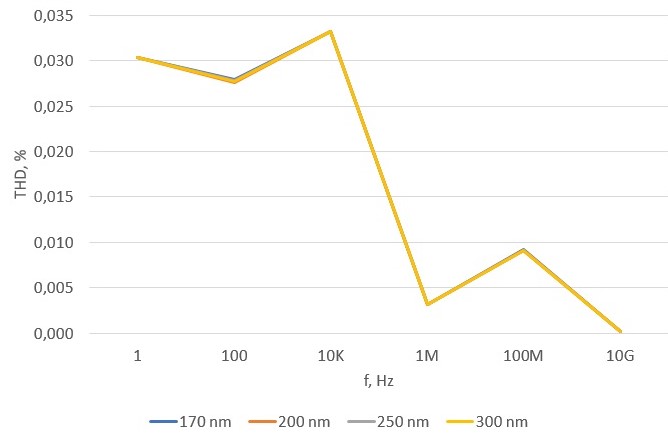} 
\caption{\small \sl THD in a CM with memristor for different channel lengths}
\label{noise_basic} 
\end{center}     
\end{figure}

\subsection{Channel length dependence}
The same procedure is done for different channel length values. The values considered were $1.7\cdot10^{-8}$, $2.0\cdot10^{-8}$, $2.5\cdot10^{-8}$ and $3.0\cdot10^{-8}$m. The resultant change in the total harmonic distortion is very small, in the order of $10^{-6}$. This change is irrespective of frequency as can be seen in the Fig. 7 and 8. The average slope for six frequencies of the CM with memristor is calculated to be 6.27$1.7\cdot10^{-8}$ per $\mu$m and for the CM without it 1.95e-5 per $\mu$m.

\begin{figure}  [!ht] 
\begin{center}  
\includegraphics[scale=0.4]{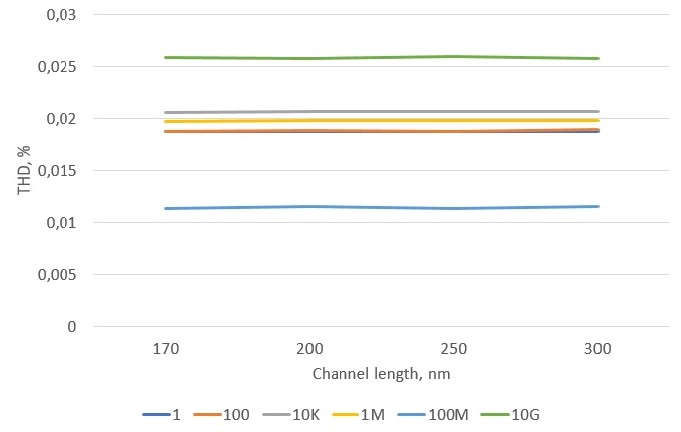} 
\caption{\small \sl THD in a CM without memristor for different frequencies}
\label{noise_basic} 
\end{center}     
\end{figure}

\begin{figure}  [!ht] 
\begin{center}  
\includegraphics[scale=0.4]{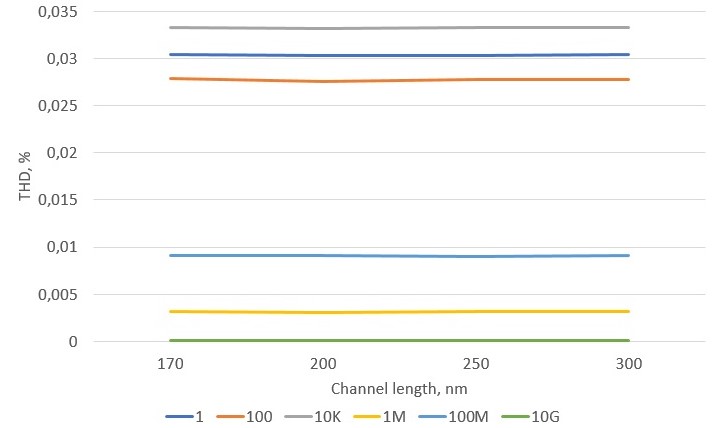} 
\caption{\small \sl THD in a CM with memristor for different frequencies}
\label{noise_basic} 
\end{center}     
\end{figure}

\subsection{Channel length dependence for the DC source}
When the channel length of the transistor which is in series with the output is changed the reference current does not change, but some discrepancy occurs between it and the output current that previously was not observed. Thus, for the source voltage amplitude equal to 3V, the current changes by 74.9$\mu $A per nm change of the length; for 5V source this value becomes 146$\mu $A per nm; and 229$\mu $A per nm for 8V supply voltage. For the basic current mirror without memristors, the current on channel length dependence is higher, namely 86$\mu $A per nm change of the length for a 3V source as shown in Fig. 9.

\begin{figure}  [!ht] 
\begin{center}  
\includegraphics[scale=0.5]{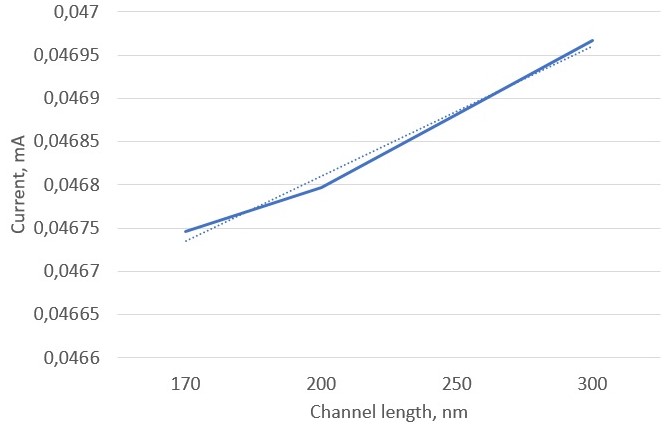} 
\caption{\small \sl Current dependence on the channel length of M2 for 3V}
\end{center}     
\end{figure}

\begin{figure}  [!ht] 
\begin{center}  
\includegraphics[scale=0.5]{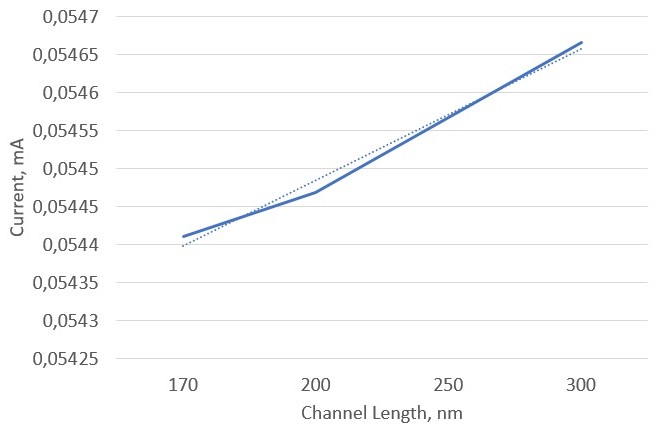} 
\caption{\small \sl Current dependence on the channel length of M2 for 3V in a basic current mirror}
\end{center}     
\end{figure}

When the length of another transistor, M1, is increased it affects both current values. In particular, the reference current increases, and the output current decreases, though much slower. Notably, the discrepancy between two currents in both cases is the same. Thus, for the 3V source voltage the difference between reference current and output current increased by 74.9$\mu $A per nm change in the length of M1 as well.

\begin{figure}  [!ht] 
\begin{center}  
\includegraphics[scale=0.5]{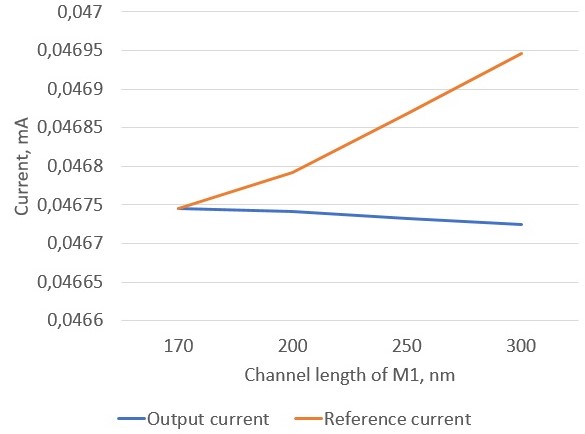} 
\caption{\small \sl Current dependence on the channel length of M1 for 3V}
\end{center}     
\end{figure}

\subsection{Memristor resistance dependence for the DC source}
Similar procedure is performed for the ON resistance of memristor. The variation is set to 10\% of the initial value. As expected, the current is decreasing as the resistance value increases linearly. For the source voltage value equal to 3V the difference between the reference and the output currents rises by 13$\mu$A\slash$\Omega$, for 5V by 1.14$\mu$A\slash$\Omega$ and for 8V by 2.67$\mu$A\slash$\Omega$.

\begin{figure}  [!ht] 
\begin{center}  
\includegraphics[scale=0.45]{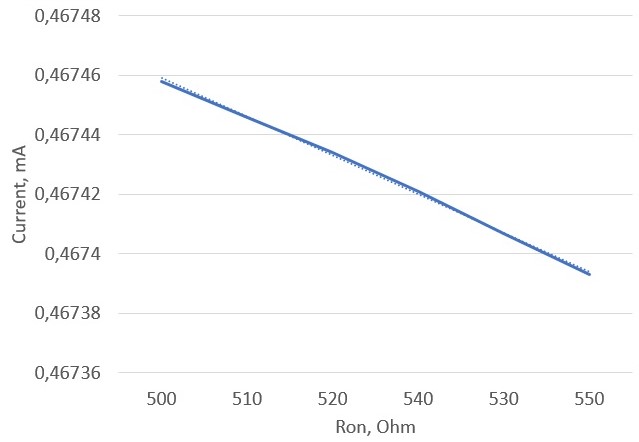} 
\caption{\small \sl Current dependence on the memristor resistance for 3V}
\end{center}     
\end{figure}

\section{Discussion}
The AC analysis results suggest that for a wide range of frequencies the CM with memristor has higher THD level than the basic CM. However, at high frequencies the THD value of the circuit with memristor goes to almost zero. In addition, the change in transistor channel length affects the proposed circuit less, which can be useful in minimizing the effect of inaccuracies that frequently occur in manufacturing process. 

The DC analysis of the circuit also shows that the discrepancy that occurs because of the channel length imbalance is less for the CM with memristor when compared to the basic CM. Also, when memristor resistance variations are considered and changed from  500$\Omega$ to 550$\Omega$ the discrepancy between the output current and the reference current is much smaller for memristor resistance variations when compared to the transistor channel length imbalance.

This analysis is limited to the comparison of the basic CMs with and without memristor. Similar comparison can be drawn between the cascode CM and a basic CM with memristor. Also, a basic CM with resistors instead of memristors could be considered. Memristor in this circuit does not really utilize its memory effect, so the main reason for its introduction into the circuit is to save area. However, in this paper the case when the voltage level of the source is close to the threshold voltage of memristor were not considered. So, other possible advantageous properties could be observed in that case. Moreover, a more complete analysis could be performed by introducing the noise of memristor and using a more practical transistor models.

\section{Conclusion}
In conclusion, the analysis of noise in the CM with memristor is presented in this paper. Both AC and DC cases are considered for transistor channel length and memristor resistance variations. The results show that memristor in the basic current mirror makes its noise level less at high frequencies, and also decreases the dependence on channel length variations significantly. Small area of memristor makes the proposed circuit a viable option when high transistor variation tolerance or low noise at high frequencies is required. The analysis, however, is limited to the comparison with the basic CM. It can be further extended by comparing the circuit with the other configurations. In addition, transistor and memristor models do not exhibit real-life properties of the devices, which  puts a limitation to the viability of results obtained.

\ifCLASSOPTIONcaptionsoff
  \newpage
\fi

\end{document}